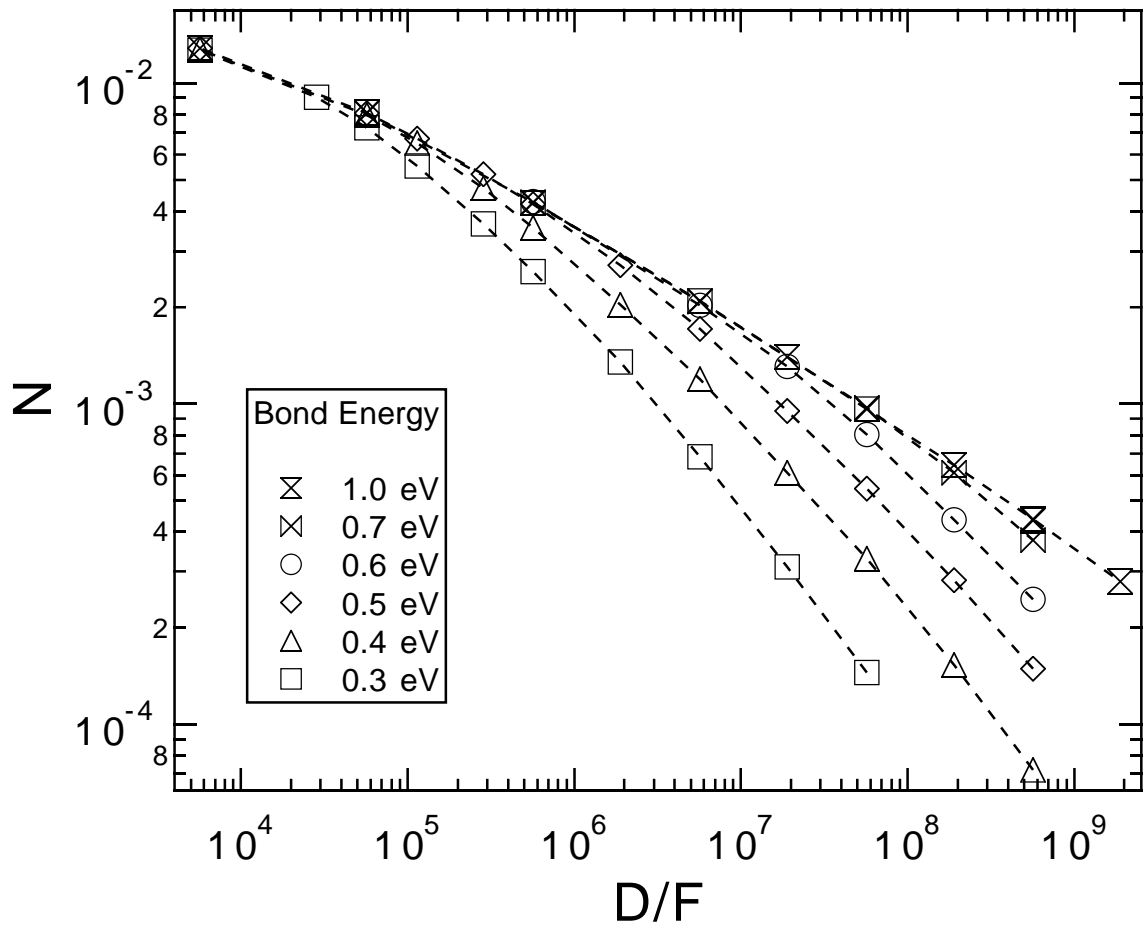

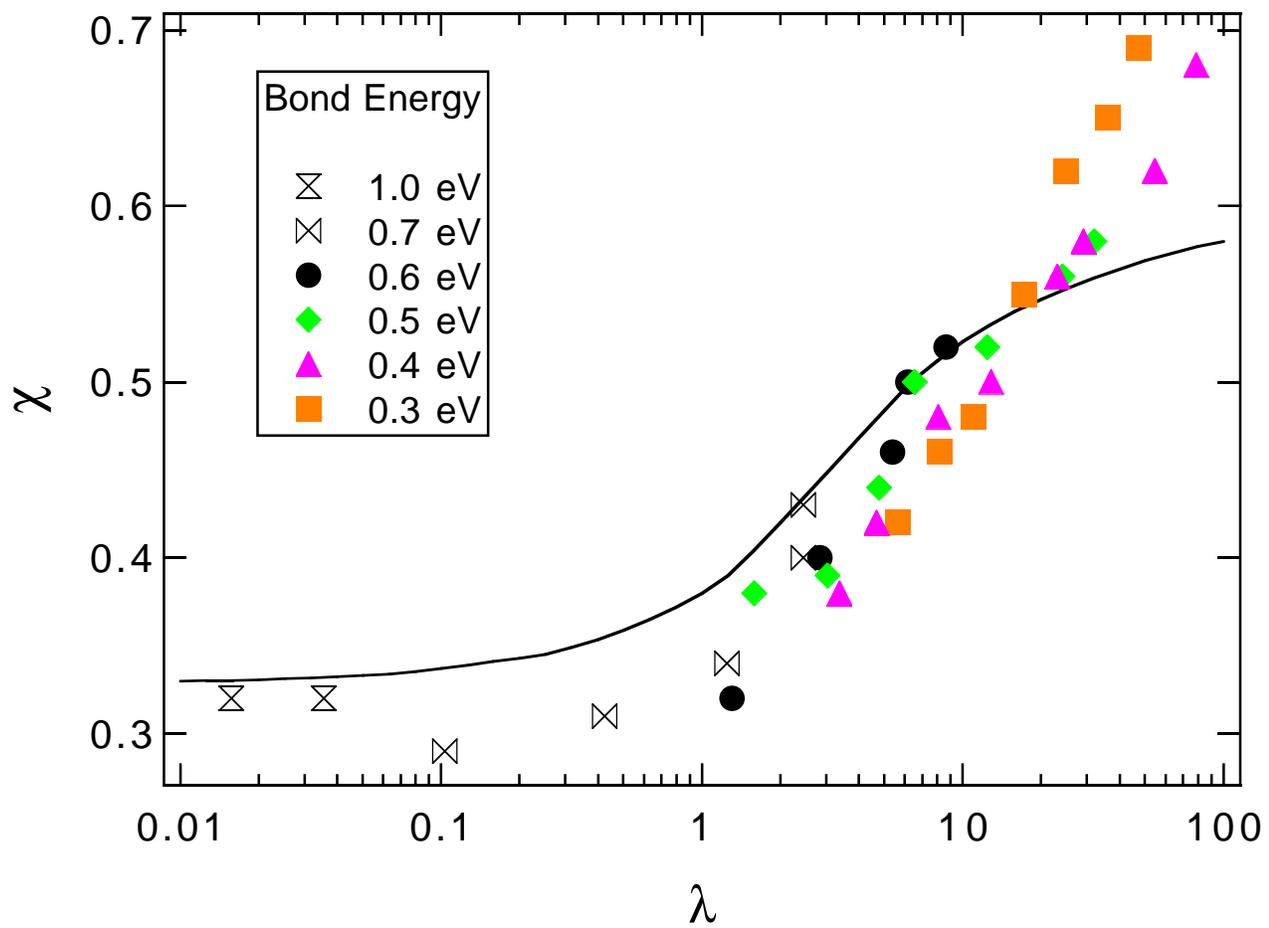

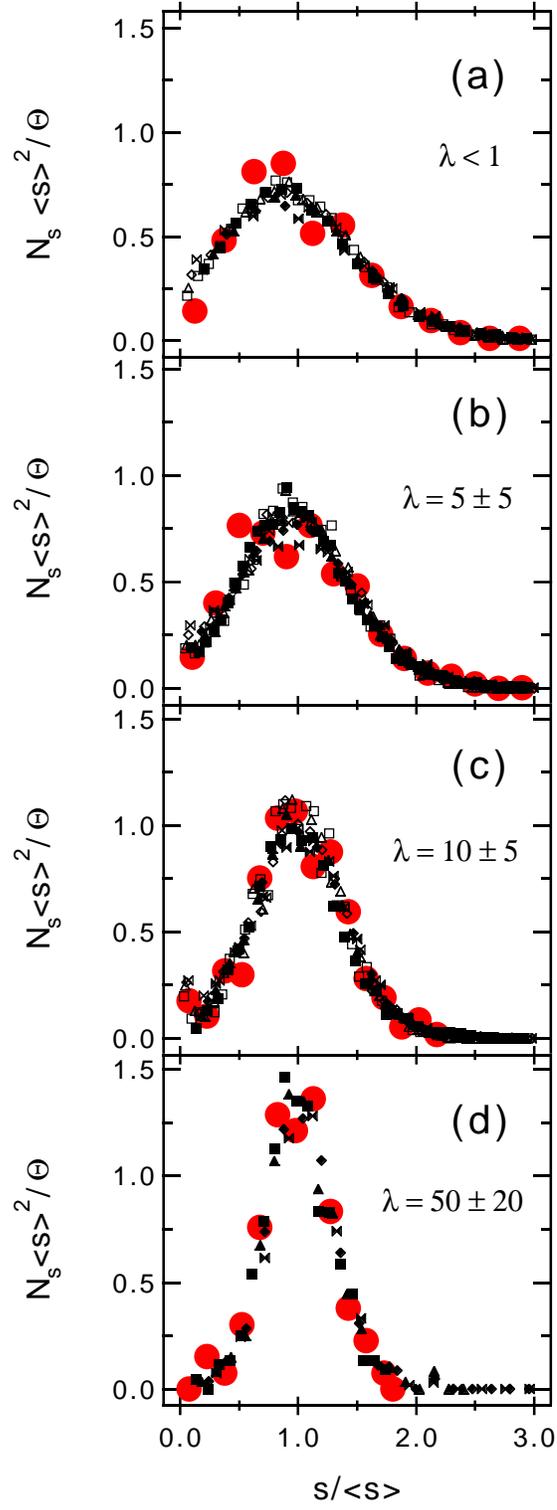

# Submonolayer Epitaxy Without A Critical Nucleus

C. Ratsch* and P. Šmilauer†

*Interdisciplinary Research Centre for Semiconductor Materials,*
*Imperial College, London SW7 2BZ, United Kingdom*

A. Zangwill

*School of Physics, Georgia Institute of Technology, Atlanta, Georgia 30332*

D.D. Vvedensky

*The Blackett Laboratory, Imperial College, London SW7 2BZ, United Kingdom*

## Abstract

The nucleation and growth of two–dimensional islands is studied with Monte Carlo simulations of a pair–bond solid–on–solid model of epitaxial growth. The conventional description of this problem in terms of a well–defined critical island size fails because no islands are absolutely stable against single atom detachment by thermal bond breaking. When two–bond scission is negligible, we find that the ratio of the dimer dissociation rate to the rate of adatom capture by dimers uniquely indexes both the island size distribution scaling function and the dependence of the island density on the flux and the substrate temperature. Effective pair-bond model parameters are found that yield excellent quantitative agreement with scaling functions measured for Fe/Fe(001).

In recent years, scanning tunnelling microscopy [1,2,3] and surface sensitive diffraction [4,5,6,7] have been applied to the study of two–dimensional nucleation and growth processes on single crystal surfaces. A substantial impetus for these studies is the fact that the standard rate equation analysis [8,9,10] of this problem implies that, if islands composed of s>i atoms do not dissociate, the total island density in the precoalescence regime varies as

$$N \sim \Theta^{1/(i+2)} \left(\frac{F}{D}\right)^\chi \exp\left[\frac{E(i)}{(i+2)k_B T}\right] \qquad \text{with} \qquad \chi = \frac{i}{i+2} \qquad (1)$$

where $\Theta$ is the coverage, F is the deposition rate, T is the substrate temperature, $E(i)$ is the cohesive energy of the most stable two–dimensional island of size $i$, $D = \nu \exp(-E_S/k_B T)$ is the single atom surface diffusion rate with attempt frequency $\nu$ and barrier $E_S$, and $k_B$ is the Boltzmann's constant. This result is of considerable importance because the microscopic parameters

$\nu$, $E_S$, and $E(i)$ are otherwise notoriously difficult to obtain [11]. On the other hand, the early careful discussion of Frankl and Venables [12] already cast doubt on the status of the so-called "critical nucleus size" $i$ defined above since " ... absolute stability does not exist for [islands] of any size. Nevertheless, the assumption that it exists for all sizes larger than the critical size is widely used, mainly because it does simplify the mathematics". These authors noted that the best case scenario occurs if "all islands quickly assume their most stable configuration".

The technique of Monte Carlo simulation is well suited for the study of many of the issues raised by experimental work and rate equation theory. Submonolayer epitaxial growth models treat the relevant physical processes—deposition, surface diffusion, aggregation, bond formation, and bond breaking—at various levels of sophistication. Examples include (i) *point island* models, where the size of growing islands is neglected [13]; (ii) *legislated $i$* models, where the critical nucleus assumption is enforced explicitly [14, 15]; (iii) *restructuring* models, where islands assume their most stable configuration either instantaneously [16] or in a controlled manner [17]; (iv) *pair-bond* models, where the energy barrier for n-coordinated atoms to hop to nearest neighbor sites is proportional to n [18,19]; (v) *restricted pair-bond* models, where the energy barrier for n-coordinated atoms to hop to nearest neighbor sites is infinite when n exceeds a specified value [20,21]; and (vi) *"realistic"* models where the rates for the elementary surface processes listed above are computed from an approximate many-body energy functional [22,23,24].

The last of these clearly is to be preferred in principle though the computation of the relevant energy functional and the identification of the pertinent microscopic processes both represent formidable challenges. For these reasons, we believe that the *pair-bond* model, where the effect of changes to the model parameters are easy to study, represents a good compromise between a realistic model and the various "toy" models with restricted dynamics. In our implementation, growth is initiated by random deposition onto a square lattice substrate at a rate F. The rate at which *any* surface atom hops to a nearest neighbor site is $\tau_n^{-1} = (2k_BT/h)\exp\left[-(E_S + nE_N)/k_BT\right] = D\exp\left(-nE_N/k_BT\right)$ where n=0,1,2,3,4 is the number of lateral nearest neighbors *before* the hop occurs and h is Planck's constant. In the simulations reported here, we have set $E_S$=1.3 eV and varied $E_N$.

In a previous study using this model [19], we varied the pair-bond energy $E_N$ by a factor of four and found the scaling in (1) for about one decade in F/D. But $\chi$ varied *continuously* in a manner that led us to suggest that the quantity $i$ was unlikely to have the direct physical meaning noted above. Subsequently, a restricted pair-bond model simulation [21] and a numerical rate equation analysis [25] also found continuous variation of $\chi$. These studies characterized this behavior as a "crossover" between well-defined regimes with $i$=1 and $i$=3.

We extend the study of Ref. [19] to a much wider range of D/F in the present work. The log-log plot in Fig. 1 shows our simulation results for the total island density versus D/F at fixed coverage $\Theta$=0.05 for various values of $E_N$. All results are for T=750 K and represent averages over at least 15 realizations on lattices of at least 500×500 sites. Spot checks with lattices of sizes $800 \times 800$ revealed that finite size effects produce changes smaller than the size of the symbols. Except possibly for $E_N$=1.0 eV, where single bond breaking is negligible [19], it is evident that straight line fits to these curves are valid only for about one decade even at the largest values of D/F. Thus, the corresponding slopes define only local, *effective* values of the exponent $\chi$. Using (1), these can be used to define effective (generally non–integer) values for $i$ [19].

Inspection of the slopes in Fig. 1 reveals that widely different choices for $E_N$ and D/F can yield the same values for $\chi$. Bartelt, Perkins and Evans [25] have suggested that when two–bond scission is negligible, the value of $\chi$ depends only on the ratio $\lambda$ of the dimer dissociation rate to the rate of adatom capture by dimers. A similar result is implicit in the formulation of Schroeder and Wolf [15]. If $N_1$ and $N_2$ denote the number density of single adatoms and immobile dimers, the rate equation estimate of this ratio is

$$\lambda = \frac{N_2/\tau_1}{DN_1N_2} = N_1^{-1} \exp\left(-E_N/k_BT\right) \tag{2}$$

where, from above, $\tau_1$ is the mean time for a dimer to dissociate [26]. A plot of $\chi$ measured from Fig. 1 versus $\lambda$ can be constructed easily because the simulation produces values for the adatom density $N_1$ for each choice of D/F [27]. Figure 2 shows that very good data collapse occurs so long as $\lambda \leq 20$ for the range of parameters studied. We have verified that two bond scission is important for those points that fail to collapse—an unsurprising fact given that $\lambda$ does not depend on $\tau_2$. Unfortunately, due in part to computational limitations, we have not been able to construct and test a generalization of $\lambda$ that collapses *all* of the data in Fig. 2.

Be that as it may, a non–trivial conclusion can be deduced by comparing our data with the rate equation prediction of Ref. [25], which is shown as a solid line in Fig. 2. By construction, the latter exhibits a smooth crossover from a plateau at $\chi = \frac{1}{3}$ ($i$=1), where dimers are absolutely stable, to a plateau at $\chi = \frac{3}{5}$ ($i$=3), where a square island of four atoms is absolutely stable [28]. Inspection of our data at fixed $E_N$ reveals a change in the sign of the *concavity* of $\chi$ versus $\lambda$ just at $\lambda \simeq 10$. This inflection point is a vestige of the $i$=3 plateau. This "plateau" would be even *less* distinct for data obtained with smaller values of $E_N$ but would be *more* distinct for simulations with larger values of $E_N$ and larger values of D/F. In fact, for *any* choice of $E_N$, the effective $\chi$ is guaranteed to deviate upward away from the solid curve for sufficiently large values of D/F (large values of $\lambda$) because the rate of two–bond scission is non–zero (see also the discussion of the upper limit of the $i$=3 plateau in [25]). A numerical study that does not artificially contract the full set of rate equations by the introduction of a critical island size undoubtedly would find the same. For this reason, no integer value of the critical island size $i$>1 is truly meaningful for this model.

In addition to total island densities, the full submonolayer island size distribution also has been the subject of recent experimental study [2, 7]. As first suggested on the basis of simulation results [13], the measurements reveal that this quantity takes the form

$$N_s = \frac{\Theta}{\langle s \rangle^2} g\big(s/\langle s \rangle\big) \tag{3}$$

where $N_s$ is the number density of islands composed of s atoms, $\langle s \rangle$ is the average island size, and g(x) is a scaling function. Till now, rate equation [25, 29] and simulation [15, 21] studies have assigned an integer index to g(x)—the presumed critical nucleus size $i$. But it is clear from the discussion above that the continuous variable $\lambda$ is a more natural choice. Figure 3 demonstrates that this is indeed the case. The circles in each panel are the scanning tunnelling microscopy data of Stroscio and Pierce for Fe/Fe(001) homoepitaxy [2, 30]. Using the bond energy extracted by these authors using (1), our simulations yield the same relative rates for atom motion with the choice $E_N \simeq 0.6$. Regrettably, most of the experimental data shown in Fig. 3 was obtained for $D/F > 4 \times 10^8$, i.e., values larger than those we could simulate with this value of $E_N$. Nonetheless, judicious choices for $\lambda$ yielded excellent fits to the experimental data over the full range of $s/\langle s \rangle$. The simulation results shown in each panel were obtained at four coverages to illustrate the data collapse predicted by (3). More importantly, each of the top three panels includes simulation data for two different choices of $D/F$ and $E_N$ that correspond to the same value of $\lambda$. But as noted, *neither* choice corresponds to the deposition conditions used in and the energy parameters extracted from the experiments. What is important is the *combination* of these factors embodied in the definition of $\lambda$.

The simulation data in the bottom panel of Figure 3 correspond to the square symbol with the largest value of $\chi$ in Figure 2 ($\chi \simeq 0.69$). But the curvature trends in Figure 1 suggest that a very similar local slope would result from the extrapolation of the $E_N \simeq 0.6$ eV curve to the value of $D/F \simeq 1.7 \times 10^{10}$ appropriate to the experimental data in that panel. Equation (1) then implies an effective critical nucleus size of about 4.5 rather than the value $i=3$ inferred from the simplified models used in Refs. [21] and [25]. The fact that our effective $i$ exceeds three implies that thermal dissociation of doubly-coordinated atoms is not negligible. The non-integer character may be interpreted as the result of a statistical average over islands of size four and larger with different shapes and relative stablity. It is worth noting that the assignment of $i=3$ in the original experimental paper [2] was made by comparing the measured value of the total island density with (1). But this formula is known [19] to have an incorrect dependence on $\Theta$ which leads to an underestimate of the total number density of islands in the present case. The foregoing extrapolation of the $E_N \simeq 0.6$ eV curve yields a number much closer to experiment. We encourage a direct measurement of $\chi$ to establish definitively whether two–bond scission is operative in this experimental range.

Two issues merit discussion at this point. First, the results shown in Figure 3 clearly imply that *effective* pair-bond parameters $E_S$ and $E_N$ can be found that reproduce the experimental aggregation data when simulations are performed using the experimental temperature and flux values. But this success does *not* imply that the pair-bond assumption used by us to model the kinetics is at all adequate as a model for the cohesion of this system. A similar situation was encountered for the case of reconstructed GaAs(001) [18] where parameters of the present model were found that provided a remarkably good account of reflection high energy electron diffraction data over a broad range of deposition conditions. On the other hand, we do feel justified to propose that the progressive shape evolution seen in Figure 3 does indeed represent a smooth thermal evolution from irreversible island growth to growth with dissocation of some singly-coordinated atoms and finally to growth with dissociation of singly and some doubly-coordinated atoms.

Second, the two–dimensional islands shown in Ref. [2] are quite compact even at T=20°C where it is well established that thermal detachment of atoms from islands is negligible. This implies that edge diffusion rapidly smoothes out an island morphology that would otherwise be fractal [17]. But for the model used here, the energy barrier to atom detachment from an island is *equal* to the energy barrier to edge diffusion [31]. The results of Figure 3 then imply that rapid edge diffusion is not essential to the form of the island size distributions found so far for Fe(001). Restricted pair bond model simulations with and without enhanced edge diffusion provide support for this assertion [21].

This is interesting because it has been suggested [25] that (1) is valid for Fe(001) but fails for our pair–bond model precisely because the lack of rapid edge diffusion precludes rapid restructuring of islands to their most compact shape. But when $E_N/k_BT$ is not very large the island shapes we obtain are in fact quite compact [19]. This is so because an atom in the simulation that detaches from an island re–attaches to that island with a high probability rather than wandering away (in contrast to the rate equation treatment when it can attach to *any* island) [32]. When $\lambda \gg 1$, islands compactify between successive aggregation events because this process repeats rapidly until a doubly bonded edge site is found. We thus hold to our view that the primary cause of the failure of (1) for the pair–bond model is simply the artificial assumption of absolute island stability needed to obtain (1).

To conclude, we remind the reader that interest in the present problem derives mostly from the hope that the rate equation result (1) facilitates the unambiguous extraction of microscopic parameters from experiment. In our view, this can be done reliably for the single–atom parameters $\nu$ and $E_S$ from a (low) temperature Arrhenius plot if, at fixed coverage and at any temperature within the plot range, *either* the scaled island size distribution has the characteristic shape shown in Fig. 3(a) *or* a regime of flux can be found where $N \sim F^{-1/3}$. Both are indicative of $i=1$. We note that generalizations for $\chi$ are available if either island diffusion or diffusional anisotropy cannot be neglected [10].

Determination of the island cohesive energy $E(i)$ is much less straightforward. Only if an extended range (*more* than one decade) of D/F is found where (1) is satisfied with an integer value of $i$ can one have reasonable confidence in the conventional rate equation prescription.

As already prefigured long ago [12], this situation does not occur with an additive pair–bond model but might be valid if E($i$) increases more rapidly than a linear function of $i$. Otherwise, a specific model must be used as an ansatz and detailed comparison made between theoretical and experimental island size distributions. But even for the simplest pair bond model, the results in Figure 3 make clear why data must be collected over the widest possible range of deposition conditions if unique model parameters are to be extracted.

In summary, we have used Monte Carlo simulations of a pair–bond model of epitaxial growth to study the nucleation and growth of two–dimensional islands on a square lattice substrate. The scaling law (1) derived from conventional rate theory cannot be valid for arbitrarily large values of D/F since there is always a non–zero rate for atom detachment from an island edge in this model. When two–bond scission is unimportant, the ratio of the dimer dissociation rate to the rate of adatom capture by dimers is found uniquely to index both the island size distribution scaling function and the dependence of the island density on the flux and the substrate temperature. A parameterization of the model that yields excellent quantitative agreement with scaling functions measured for Fe/Fe(001) suggests that thermal dissociation of doubly-coordinated atoms is not negligible at the highest temperatures studied in the experiment.

We thank Jim Evans, Peter Feibelman, and Jerry Tersoff for helpful discussions and the authors of Ref. [2] for permission to show some previously unpublished data for Fe/Fe(001). Work at Georgia Tech was performed with support from the U.S. Department of Energy under Grant No. DE–FG05–88ER45369. Work at Imperial College was performed with support from a NATO travel grant and the Research Development Corporation of Japan.

## FIGURE CAPTIONS

**Figure 1.** Island density N as a function of D/F for different values of the lateral pair–bond energy $E_N$. The fixed parameters are $\Theta=0.05$, T=750 K, and $E_S=1.3$ eV.

**Figure 2.** The local slope $\chi$ as a function of $\lambda = N_1^{-1}\exp(-E_N/k_BT)$ for different values of $E_N$. $N_1$ is the measured density of adatoms. The solid curve is the rate equation result from Ref. [25].

**Figure 3.** Island size distributions for Fe/Fe(001) from Ref. [2,30] (large circles) compared to pair–bond model simulation results: (a) experiment at T=20°C, simulation for $0.10 \leq \Theta \leq 0.25$ with $E_N=1.0$ eV and F=0.1 $s^{-1}$ (open symbols) and $E_N=0.3$ eV and F=0.5 $s^{-1}$ (closed symbols); (b) experiment at T=163°C, simulation for $0.10 \leq \Theta \leq 0.25$ with $E_N=0.7$ eV and F=0.001 $s^{-1}$ (open symbols) and $E_N=0.3$ eV and F=0.2 $s^{-1}$ (closed symbols); (c) experiment at T=250°C, simulation for $0.12 \leq \Theta \leq 0.25$ with $E_N=0.6$ eV and F=0.001 $s^{-1}$ (open symbols) and $E_N=0.3$ eV and F=0.03 $s^{-1}$ (closed symbols); (d) experiment at T=356°C, simulation for $0.12 \leq \Theta \leq 0.25$ with $E_N=0.3$ eV and F=0.001 $s^{-1}$.